\documentclass[superscriptaddress,twocolumn,showpacs,prc,aps,amssymb,amsfonts,nofootinbib]{revtex4}

\usepackage{graphicx}

\begin{document}
\title{Additivity of effective quadrupole moments and angular momentum alignments
in the $A\sim130$ nuclei}
\author{M.\ Matev}
\affiliation{Department of Physics and Astronomy, University of Tennessee,
Knoxville, Tennessee 37996, USA}
\affiliation{Institute for Nuclear Research and Nuclear Energy,
Bulgarian Academy of Sciences,
72 Tzarigradsko chaussee Blvd., BG-1784, Sofia, Bulgaria
}
\author{A.\ V.\ Afanasjev}
\affiliation{Department of Physics and Astronomy, Mississippi State
University, Mississippi State, Mississippi 39762, USA}
\affiliation{Laboratory of Radiation Physics, Institute of Solid State Physics,
     University of Latvia, LV 2169, Salaspils, Miera str. 31, Latvia}

\author{J.\ Dobaczewski}
\affiliation{Department of Physics and Astronomy, University of Tennessee,
Knoxville, Tennessee 37996, USA}
\affiliation{Physics Division, Oak Ridge National Laboratory, P.O.
Box 2008, Oak Ridge, Tennessee 37831, USA}
\affiliation{Institute of Theoretical Physics, University of Warsaw, ul.
Ho\.za 69, PL-00-681 Warsaw, Poland}
\affiliation{Department of Physics, P.O. Box 35 (YFL),
FI-40014 University of Jyv\"askyl\"a, Finland}
\author{G.\ A.\ Lalazissis}
\affiliation{Department of Theoretical Physics, Aristotle University
of Thessaloniki, GR-54124, Greece}
\author{W.\ Nazarewicz}
\affiliation{Department of Physics and Astronomy, University of Tennessee,
Knoxville, Tennessee 37996, USA}
\affiliation{Physics Division, Oak Ridge National Laboratory, P.O.
Box 2008, Oak Ridge, Tennessee 37831, USA}
\affiliation{Institute of Theoretical Physics, University of Warsaw, ul.
Ho\.za 69, 00-681 Warsaw, Poland}

\date{\today}

\begin{abstract}
The additivity principle of the extreme
shell model stipulates that an average value of a
one-body operator be equal to the sum of the
core contribution and effective contributions of valence (particle or hole) nucleons.
For quadrupole moment and angular momentum operators,
 we test this
principle for highly and superdeformed rotational bands in the
$A\sim130$ nuclei. Calculations are done in the self-consistent
cranked non-relativistic Hartree-Fock
and  relativistic  Hartree
mean-field approaches.
Results indicate
that the additivity principle is a valid concept that justifies the
use of an extreme single-particle model in an unpaired regime typical
of high angular momenta.
\end{abstract}

\maketitle

\section{Introduction}
\label{intro}

The behavior of the nucleus at high angular momenta is strongly affected
by the single-particle (s.p.)  structure, i.e., shell effects. Properties of the
s.p.\ orbits around the Fermi level determine the deformability of the nucleus,
the amount of angular momentum available in the lowest-energy
configurations, the moment of inertia, and the Coriolis coupling. Consequently,
nucleonic shells can be seen and probed through the measured properties of
rapidly rotating nuclei.

The independent particle model is a  first approximation to the nuclear
motion. Here, the nucleons are assumed to move independently of each
other in an average field generated by  other nucleons. Each nucleon
occupies a s.p.\ energy level, and levels with similar energies are
bunched together into shells. The wave function of a given many-body
configuration uniquely characterized by s.p.\ occupations is an
antisymmetrized product of one-particle orbitals (the Slater
determinant). In the next step, the residual interaction between
particles needs to be considered. This is the essence of the
configuration interaction method or the interacting shell model. For
heavier nuclei, where the number of s.p.\ orbits becomes large, a
customary approximation is to divide the configuration space into the
(inert) core states and the  (active) valence orbits and to perform
configuration mixing in the valence subspace.

The basic idea behind the additivity principle for one-body operators is
rooted in the independent particle model. The principle states that the
average value of a one-body operator $\hat{O}$ in a given many-body
configuration $k$, $O(k)$, relative to the average value in the core
configuration $O^{\text{core}}$, is equal to the sum of effective
contributions of particle and hole states by which the $k$-th
configuration differs from that of the core. Such a property is
trivially valid in the independent particle model. However, the presence
of residual interactions and resulting configuration mixing could, in principle,
spoil the simple picture. In particular, in the interacting shell
model, the polarization effects due to additions of particles or holes
are significant and they give rise to strong modifications of the mean field.
So the essence of the additivity
principle lies in the fact that these polarizations are, to a large
extent, independent of one another and thus can by treated additively.

The additivity principle for strongly deformed nuclear systems was emerging
gradually  in the 1990s. First, it was found in Ref.\ \cite{R.91}
that effective (relative) angular momentum
alignments are additive to a good precision in the
superdeformed (SD) bands around $^{147}$Gd. However, the analysis was only
restricted to a few bands. Later, the statistical analysis of
Ref.\ \cite{FBHN.96} in
the $A\sim 150$ and 190 mass regions clearly demonstrated that the so-called
phenomenon of band twinning (or identical bands) is more likely to occur in
SD than in normal-deformed bands. It was shown that a necessary condition
for the occurrence of identical bands is the presence of the same number of  high-$N$
intruder orbitals (see also Ref.\ \cite{KRA.98}). In addition, it was
concluded that the
configuration-mixing interactions such as pairing and the coupling to the low-lying
collective vibrational degrees of freedom act destructively on identical bands by
smearing out the individuality of each s.p.\ orbital. Such individuality
is an important ingredient for the
additivity principle:
it is expected that this principle works only in the systems with weak residual
interaction, in particular, pairing \cite{FBHN.96,ZSRG.98}.

The principle of additivity at superdeformation was explicitly and
thoroughly formulated for the case of the $Q_{20}$ quadrupole moments in
the non-relativistic study of quadrupole moments of SD bands in the
$A\sim 150$ mass region in Ref.\ \cite{SDDN.96} within the
cranked Hartree-Fock (CHF) approach based on Skyrme forces. It was
shown that the charge quadrupole moments calculated with respect to the
doubly magic SD core of $^{152}$Dy can be expressed very precisely in
terms of effective contributions from the individual hole and particle
orbitals, independently of the intrinsic configuration and of the
combination of proton and neutron numbers.

Following this work, it was
shown that the principle of additivity of quadrupole moments works also
in the framework of the  microscopic+macroscopic method (in particular,
the configuration-dependent cranked Nilsson+Strutinsky approach)
\cite{AJKR.97,KRA.98np}. However, contrary to self-consistent
approaches, the effective s.p.\ quadrupole moments of the
microscopic+macroscopic method are not uniquely defined due to the lack
of self-consistency between the microscopic and macroscopic
contributions.

The study of additivity of quadrupole moments and
effective alignments was also performed in the framework of the cranked
relativistic mean field (CRMF) approach, but it was restricted to a few
configurations in the vicinity of the doubly  magic SD core of $^{152}$Dy
\cite{ALR.98}. It was suggested in this work that the additivity
principle when 
applied to the angular momentum operator (i.e., effective
alignments) does not work as well as for the quadrupole moment. In
addition, the effective alignments of high-$N$ intruder orbitals seem to
be  less additive than  the effective alignments of non-intruder
orbitals. The latter can be attributed to a pronounced polarization of
the nucleus by high-$N$ intruder orbitals at high spin.

\begin{figure*}
\centering
\includegraphics[width=13cm]{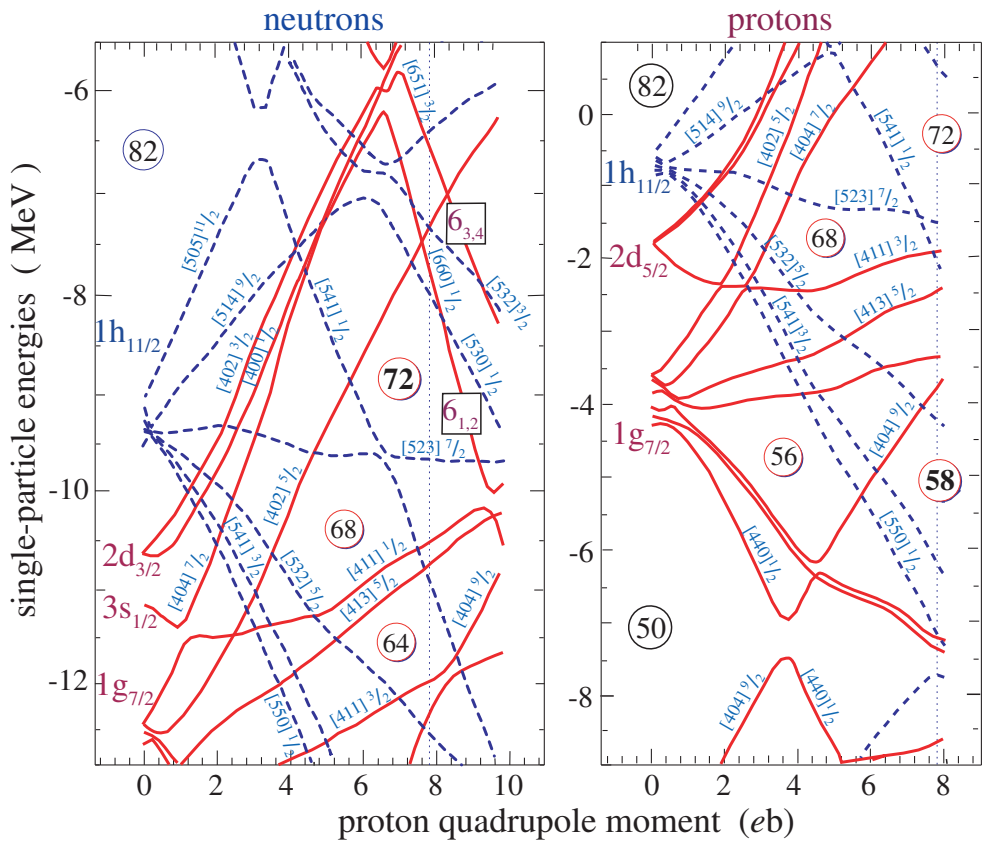}
\caption{(color online) Single-particle energies for neutrons (left) and protons
(right)
in $^{128}$Ba as a function of the
proton quadrupole moment
calculated in the HF+SLy4  model.
Solid and dashed lines mark positive
and negative parity states, respectively. The orbitals are labeled by
the asymptotic (Nilsson)
quantum numbers $[{\cal N}n_z\Lambda]\Omega$
of the dominant
component of the s.p.\ wave function.
The neutron intruder orbitals originating from the
${\cal N}$=6 shell  are additionally labeled by
the main oscillator quantum
number  and a subscript denoting the position of the
orbital within the ${\cal N}$ shell.}
\label{Nils-diag-Ba128}
\end{figure*}

For quadrupole moments, the additivity principle was experimentally
confirmed in the $A$$\sim$140-150 mass region of superdeformation. It
was shown that the quadrupole moments of the SD bands in $^{142}$Sm
\cite{142Sm-exp} and $^{146}$Gd \cite{146Gd-exp} could be  well
explained in terms of the $^{152}$Dy SD core  and effective
s.p.\ quadrupole moments of valence (particle and hole)
orbits. All of these studies, together with the previous results for
moments of inertia \cite{NWJ.89,AKR.96} and effective alignments
\cite{Rag.93,ALR.98}, strongly suggest that the SD  bands in
the $A$$\sim$140-150 mass region are excellent examples of an almost
undisturbed s.p.\ motion. This is especially true at
rotational frequencies above $\hbar\omega$=0.5 MeV \cite{AKR.96,ALR.98} where pairing
is expected to be  of minor importance. (For other excellent  examples
of an almost undisturbed s.p.\ motion at high spins, see
Refs.~\cite{AFLR.99,SW.05,Sto.06}.)

In the mass $A$$\sim$135\,\,\,($Z$=58-62) light rare-earth region,
large $Z$=58 and $N$=72 shell gaps (see Fig.\ \ref{Nils-diag-Ba128} and
Refs.\ \cite{Wyss.88,AR.96}) lead to the existence of rotational
structures with characteristics typical of highly deformed and SD
bands. These bands were observed up to high and very high spins (see
Refs.\ \cite{Ce132-131,A130-exp2} and references quoted therein). For
example, the yrast SD band in $^{132}$Ce extends  to $\sim$68$\hbar$,
which represents one of the highest spin states ever observed in atomic
nuclei \cite{Ce132-131}. At such high spins, pairing is expected to
play a minor role \cite{NWJ.89,AR.96,A130-exp1}, which is a necessary condition
for the additivity principle to hold. In this mass region, experimental
studies of the additivity principle were performed in Refs.\
\cite{A130-exp1,A130-exp2}. Differential lifetime measurements, free
from common systematic errors, were performed for over 15 different
nuclei (various isotopes of Ce, Pr, Nd, Pm, and Sm) at high spin within
a single experiment \cite{A130-exp1,A130-exp2}.

There are several notable differences between the $A$$\sim$135 and
$A$$\sim$140-150 regions of superdeformation. In particular, the
rotational bands in the $A$$\sim$135 region are calculated to correspond
to the local energy minima  that are characterized by much larger
$\gamma$-softness than those in the $A$$\sim$140-150 mass region
\cite{Wyss.88,AR.96}. Thus, one of the main goals of the present
manuscript is to find the impact of the $\gamma$-softness on the
additivity principle. The second goal is a detailed study of the
additivity principle not only for quadrupole moments but also for
angular momentum  alignments. The present work is the first study where
the additivity of relative alignments has been tested within the CHF and
CRMF frameworks in a systematic way along with the additivity of
quadrupole moments.  Some results of this study have been  reported in
Refs.\ \cite{A130-exp1,A130-exp2}.

This paper is organized as follows. The principle of
additivity, definitions of physical observables, the way of finding
effective s.p.\ quantities, and details of theoretical
calculations are discussed in Sec.\ \ref{Principle-of-additivy}.
Analysis of the additivity principle for quadrupole moments and relative
alignments, and the discussion of associated theoretical  uncertainties
are presented in Sec.\ \ref{res-add}.
Finally,  Sec.\ \ref{Concl} contains the  main conclusions of our work.

\begin{figure}
\centering
\includegraphics[width=6cm]{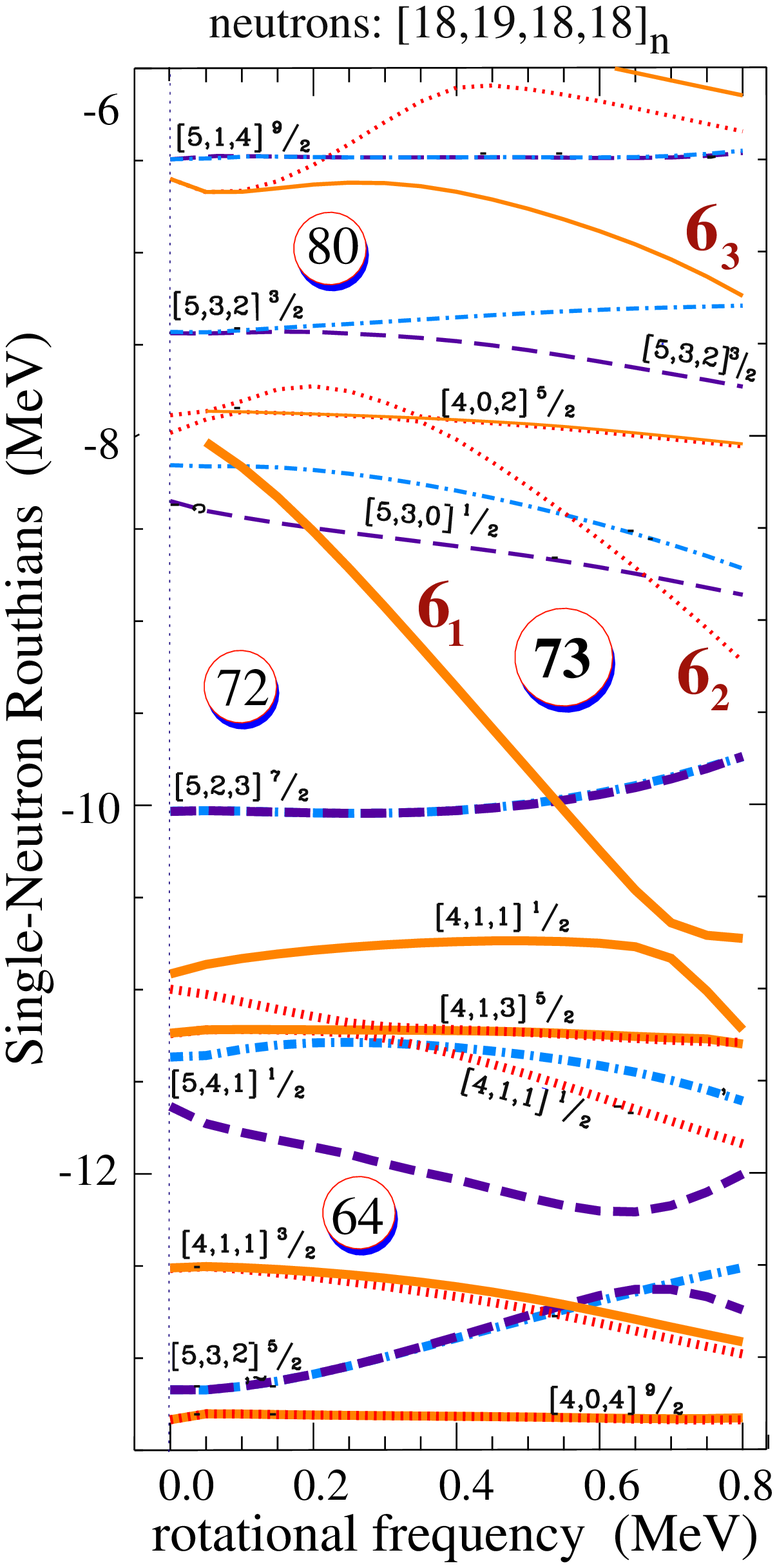}
\caption{(color online) Neutron s.p.\ energies (routhians)
 in the self-consistent
rotating potential (CHF+SLy4) as a function of rotational frequency for the core
configuration (the lowest highly deformed band of $^{131}$Ce).
 Occupied (empty) states are denoted
by thick (thin) lines.
Dotted, solid,
dot-dashed, and dashed lines indicate parity and signature quantum
numbers $(\pi,r)$=$(+,+i)$, $(+,-i)$, ($-,+i)$, and $(-,-i)$, respectively.
The orbitals are also labeled by the asymptotic quantum
numbers $[{\cal N}n_z\Lambda]\Omega$  of the dominant harmonic oscillator
component of the wave function.
The neutron intruder orbitals originating from the
${\cal N}$=6 shell  are additionally marked. At intermediate rotational frequencies,
the lowest intruder level 6$_1$ becomes occupied and this leads to the presence
of the large gap in the spectrum at $N$=73.
}
\label{131Ce-sp-neut}
\end{figure}
\begin{figure}
\centering
\includegraphics[width=6cm]{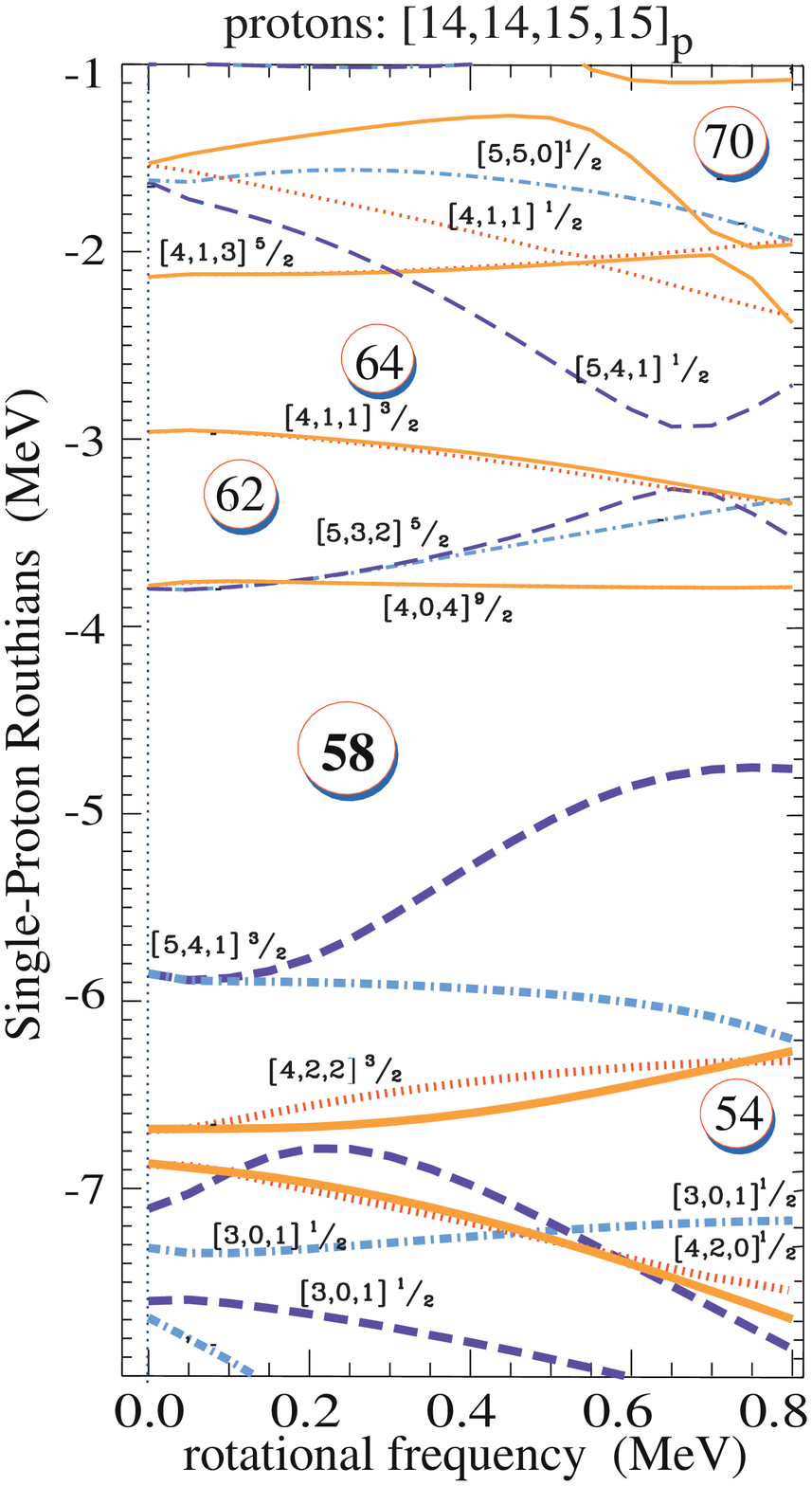}
\caption{(color online) Similar to Fig.\ \protect\ref{131Ce-sp-neut}
except  for proton
s.p.\ states. The large proton $Z$=58 gap in the s.p.\ spectrum
is present at all frequencies considered.
\label{131Ce-sp-prot}}
\end{figure}

\section{Theoretical framework}
\label{Principle-of-additivy}

\subsection{Definition of observables}
\label{observables}

Since  pairing is neglected in this work, the charge quadrupole moments
$Q_{20}$ and
$Q_{22}$ are defined microscopically as sums of
expectation values of the s.p.\ quadrupole moment operators
$\hat{q}_{20}$ and $\hat{q}_{22}$ of the occupied proton states, i.e.,
\begin{eqnarray}
\label{Qd-a}
Q_{20} &=& \sum_\mu \left< \mu | \hat{q}_{20} | \mu \right>, \\
Q_{22} &=& \sum_\mu \left< \mu | \hat{q}_{22} | \mu \right>,
\label{Qd-b}
\end{eqnarray}
where $\hat{q}_{20}$ and $\hat{q}_{22}$ are defined in three-dimensional
Cartesian coordinates as \cite{Qdef} (conserved signature symmetry is assumed)
\begin{eqnarray}
\hat{q}_{20} &=& 2z^2-x^2-y^2, \\
\hat{q}_{22} &=& \sqrt{3}(x^2-y^2).
\end{eqnarray}
The factor of $\sqrt{3}$ is included in the definition of $\hat{q}_{22}$
in order to have the following expressions for the total quadrupole moment
$Q_2$ and the associated Bohr angle $\gamma$:
\begin{eqnarray}
Q_2          &=& \sqrt{Q_{20}^2+Q_{22}^2}, \\
\tan(\gamma) &=& Q_{22}/Q_{20}. \label{def-gamma}
\end{eqnarray}

Note that the sums in Eqs.\ (\ref{Qd-a}--\ref{Qd-b}) run only over
proton states. The neutrons, having zero electric charge, do not appear in the
sums explicitly, but they influence
the charge quadrupole moments  indirectly via the
quadrupole polarization
(deformation changes) induced by  occupying/emptying single-neutron states.

It should be noted that with the definitions (\ref{Qd-a}--\ref{def-gamma}),
the spherical components of the quadrupole tensor are $Q_{20}$ and
$Q_{22}/\sqrt{2}$. This fact is important for the definition of the
so-called transition quadrupole moment $Q_t$ \cite{Retal.82,[Ham86a]}.
This moment gives the measure of the transition strength of the
$\Delta I$=2 (stretched)
$E2$ radiation in the limit of large deformation and angular momentum, and it is
proportional to the component $Q^{\omega}_{22}/\sqrt{2}$ of the
spherical quadrupole tensor when the quantization axis coincides with
the vector of rotational velocity $\omega$, i.e.,
\begin{widetext}
\begin{eqnarray}
Q^{\omega}_{20} &=&
              D^2_{0, 0}(\psi^{\omega},\theta^{\omega},\phi^{\omega})Q_{20}
      + \left[D^2_{0, 2}(\psi^{\omega},\theta^{\omega},\phi^{\omega})
            + D^2_{0,-2}(\psi^{\omega},\theta^{\omega},\phi^{\omega})\right]
       \frac{Q_{22}}{\sqrt{2}} , \\
\frac{Q^{\omega}_{22}}{\sqrt{2}} &=&
              D^2_{2, 0}(\psi^{\omega},\theta^{\omega},\phi^{\omega})Q_{20}
      + \left[D^2_{2, 2}(\psi^{\omega},\theta^{\omega},\phi^{\omega})
            + D^2_{2,-2}(\psi^{\omega},\theta^{\omega},\phi^{\omega})\right]
       \frac{Q_{22}}{\sqrt{2}}.
\end{eqnarray}
\end{widetext}
Here, symbols $D^\lambda_{\mu\nu}$ denote the Wigner functions \cite{[Var88]},
with their arguments $\psi^{\omega},\theta^{\omega},\phi^{\omega}$ being
the Euler angles that rotate  the $z$ axis (the standard
quantization axis for spherical tensors) onto the direction of the
angular velocity.

For the cranking axis coinciding with the $y$-axis of the intrinsic
system, as is the case for the code HFODD \cite{DD.97a,DD.97b} used
in the present study, the Euler angles are $\psi$=0, $\theta$=$\pi/2$,
and $\phi$=$\pi/2$, which gives:
\begin{eqnarray}
      Q^{\omega\parallel{y}}_{20}            & = &
-{1\over 2}Q_{20} - \sqrt{{3\over 2}}\frac{Q_{22}}{\sqrt{2}}, \\
\frac{Q^{\omega\parallel{y}}_{22}}{\sqrt{2}} & = &
  \sqrt{3\over 8}Q_{20}  -{1\over 2}\frac{Q_{22}}{\sqrt{2}}.
\end{eqnarray}
The second of these equations  gives the definition of
the transition quadrupole moment used in this work:
\begin{equation}
Q^{\omega\parallel{y}}_{t} =
      \sqrt{8\over 3}\frac{Q^{\omega\parallel{y}}_{22}}{\sqrt{2}} =
      Q_{20} - \sqrt{{2\over3}} \frac{Q_{22}}{\sqrt{2}}.
\label{qqdef}
\end{equation}

In order to provide a link to studies that employ
the $x$-axis cranking, like, e.g.,
Refs.\ \cite{Retal.82,[Ham86a]}
and  our earlier papers \cite{A130-exp1,A130-exp2}, we repeat
derivations for the Euler angles
$\psi$=$\pi/2$, $\theta$=$\pi/2$, $\phi$=$\pi$, which rotate  the $z$ axis
onto the $x$ axis:
\begin{eqnarray}
      Q^{\omega\parallel{x}}_{20}            & = &
-{1\over 2}Q_{20} + \sqrt{{3\over 2}}\frac{Q_{22}}{\sqrt{2}}, \\
\frac{Q^{\omega\parallel{x}}_{22}}{\sqrt{2}} & = &
  -\sqrt{3\over 8}Q_{20}  -{1\over 2}\frac{Q_{22}}{\sqrt{2}},
\end{eqnarray}
hence
\begin{equation}
Q^{\omega\parallel{x}}_{t} =
      -\sqrt{8\over 3}\frac{Q^{\omega\parallel{y}}_{22}}{\sqrt{2}} =
      Q_{20} + \sqrt{{2\over3}} \frac{Q_{22}}{\sqrt{2}}.
\label{qqdefx}
\end{equation}

Although definitions (\ref{qqdef}) and (\ref{qqdefx}) differ by signs
of the second terms, values of $Q^{\omega\parallel{y}}_{t}$
and $Q^{\omega\parallel{x}}_{t}$ obtained in self-consistent
calculations must be identical because they cannot
depend on the direction of the cranking axis. It means that 
values of $Q_{22}$ obtained in cranking calculations along the $y$
and $x$ axes  have opposite signs. In what follows,
we employ definition (\ref{qqdef}) of the transition moment
and drop the superscripts that denote the direction of the
cranking axis, e.g.,
\begin{eqnarray}
\label{Qteff}
Q_{t}       &=&        Q_{20} - \sqrt{{1\over3}}\,       Q_{22}, \\
\hat{q}_{t} &=&  \hat{q}_{20} - \sqrt{{1\over3}}\, \hat{q}_{22}.
\label{qteff}
\end{eqnarray}

Finally, the expectation value of the total angular momentum 
$J$ (its projection
on the cranking axis) is defined as a sum
of the expectation values of the s.p.\ angular momentum operators
$\hat{j}_y$ of the occupied states
\begin{eqnarray}
J\equiv \langle \hat{J}_y\rangle = \sum_\mu \left< \mu | \hat{j}_y | \mu \right> .
\end{eqnarray}
The value of  $J$ can be  expressed in terms of
the total spin $I$ via the cranking relation
 \cite{I.54}
\begin{eqnarray}
J=\sqrt {I(I+1)} \approx I+\frac 12.
\end{eqnarray}

\subsection{Additivity of effective s.p.\ observables}
\label{add-prin-math}

For
each $k$-configuration
defined by occupying a given set of
s.p.\ orbitals and represented by a product state $|k\rangle$,
we determine the average value
$O(k)=\langle{k}|\hat{O}|k\rangle$ of a s.p.\ operator
$\hat{O}$. We may now  designate one of these configurations as
a reference, or a core configuration, and determine the relative
change $\delta O(k) \equiv O(k) - Q^{\text{core}}$ of the physical observable in
the $k$-th configuration with respect to that in the core
configuration. The additivity principle stipulates that all these
differences can be expressed as sums of individual effective
contributions $o^{\text{eff}}_{\alpha}$ coming from s.p.\ states
(enumerated by index $\alpha$),
i.e.,
\begin{equation}
O(k) - Q^{\text{core}} \equiv \delta O(k) = \sum_{\alpha}c_{\alpha}(k)o^{\text{eff}}_{\alpha}.
\label{add}
\end{equation}
Coefficients $c_{\alpha}$ in Eq.~(\ref{add})
 define the s.p.\ content
of the configuration $k$ with respect to the core configuration. Namely,
\begin{itemize}
\item[(i)]
$c_{\alpha}(k)=0$ if the state $\alpha$ is not occupied in either of these
two configurations, or is occupied in both of them,
\item[(ii)]
$c_{\alpha}(k)=1$ if $\alpha$ has a {\em particle} character
 (it is occupied in the $k$-th configuration and is not occupied
in the core configuration),
\item[(iii)]
$c_{\alpha}(k)=-1$ if the state $\alpha$ has a {\em hole}
character  (it is not occupied in the $k$-th
configuration and is occupied in the core configuration).
\end{itemize}
In this way, one can label the $k$-th configuration with the set of
coefficients $c(k) = \{c_{\alpha}(k), \alpha = 1, \dots ,m\}$, where
$m$ denotes the size of s.p.\ space considered. The
values of $o^{\text{eff}}_{\alpha}$ can be calculated by proceeding step by
step from the core configuration to the configurations differing by
one particle or one hole, then to the configurations differing by two
particles, two holes, or a particle and a hole,  and so forth, until the
data set is generated which is statistically large enough to provide
appreciable precision for $o_\alpha^{\text{eff}}$. Had the additivity principle been
obeyed exactly, calculations limited to  one-particle and one-hole configurations
would have sufficed. Since our goal is  not only to determine values
of $o^{\text{eff}}_{\alpha}$ but actually prove that the additivity principle
holds up to a given accuracy, we have to consider a large set of
configurations and determine the best values of $o^{\text{eff}}_{\alpha}$
together with their error bars.

In what follows, we consider relative changes in the average quadrupole
moments $\delta Q_{20}(k)$ and $\delta Q_{22}(k)$, transition
quadrupole moments $\delta Q_t(k)$, and total angular  momenta  $\delta J(k)$ (see
Sec.\ \ref{observables}), which are related to the
effective one-body expectation values  via the additivity principle.

The addition of particle or hole in a specific single-particle
orbital $\alpha$ gives rise to a polarization of the  system, so the effective
s.p.\ values, $o^{\text{eff}}_{\alpha}$, depend  not
only on  the bare s.p.\ expectation  values,
$o^{\text{bare}}_{\alpha}=\left< \hat{o}\right>_{\alpha}$, but also
contain   polarization contributions. For example, the effective
s.p.\ charge quadrupole moment $q_{20,\alpha}^{\text{eff}}$
can be represented as the sums of  bare s.p.\  charge
quadrupole moments $q_{20,\alpha}^{\text{bare}}=\left< \hat{q}_{20}
\right>_{\alpha}$ and  polarization contributions
$q^{\text{pol}}_{20,\alpha}$:
\begin{equation}
\label{def_qeff20barepol}
q^{\text{eff}}_{20,\alpha} =
  q^{\text{bare}}_{20,\alpha} + q^{\text{pol}}_{20,\alpha} .
\end{equation}
Therefore, for neutron orbitals, which have vanishing bare
charge quadrupole moments, $q^{\text{bare}}_{20,\alpha_n}=0$,
the effective charge quadrupole moments are solely given
by polarization terms:
\begin{equation}
\label{def_qeffnbarepol}
q^{\text{eff}}_{20,\alpha_n}   =  q^{\text{\text{pol}}}_{20,\alpha_n} .
\end{equation}

\subsection{Determination of effective s.p.\  observables}
\label{effect}

Once the averages of physical observables $O(k)$ for the set of $N_c$
calculated configurations ($k=1,\dots,N_c$) are determined, the
effective s.p.\  contributions $o^{\text{eff}}_{\alpha}$
(\ref{add}) are found by means of a multivariate least-square fit
(see, e.g., Refs.\ \cite{P-book,LH-book}). This is done by minimizing
the function of $o^{\text{eff}}_{\alpha}$ defined by

\begin{equation}
F\left[o^{\text{eff}}_{\alpha}\right]=\sum_{k=1}^{N_c}
\left( \delta O(k)- \sum_{\alpha=1}^m o_{\alpha}^{\text{eff}} c_{\alpha}(k) \right )^2 .
\end{equation}
Note that the problem is only meaningful when the number of configurations
 is sufficiently large, $N_c > m$. Following
the general concept of the least-square method,
the partial differentiation with respect to the variables
$o^{\text{eff}}_{\alpha}$ yields
\begin{eqnarray}
0  & = & \frac{1}{2} \frac{\partial}{\partial o_{\alpha}^{\text{eff}}}
F\left[o^{\text{eff}}_{\alpha}\right] = {}  \nonumber \\
&  = & \sum_{\alpha'} \sum_{k} c_{\alpha}(k) c_{\alpha'}(k) o^{\text{eff}}_{\alpha'}
      -\sum_{k} \delta O(k) c_{\alpha}(k)= \nonumber \\
{} & = & (Bo^{\text{eff}}-a)_{\alpha} ,
\end{eqnarray}
where $a_{\alpha}=\sum_k \delta O(k) c_{\alpha}(k) = c^T \delta O$ and $B=||B_{\alpha \alpha'}||=
||\sum_k c_{\alpha}(k) c_{\alpha'}(k) ||=c^Tc$. Solving this equation by inverting the
non-singular matrix $B$ gives the solution to the multivariate regression problem:
\begin{equation}
\tilde{o}^{\text{eff}} = B^{-1}a=(c^T c)^{-1} c^T \delta O.
\end{equation}
The fact that $B$ is positive-definite guarantees that the solution $\tilde{o}^{\text{eff}}$
corresponds to a minimum of $F\left[o^{\text{eff}}_{\alpha}\right]$.

In order to estimate the
variance, we assume that the first statistical moments of residuals,
\begin{equation}
\label{residuals}
\epsilon_O(k) = \delta O(k) - \sum_{\alpha} c_{\alpha}(k) \tilde{o}_{\alpha}^{\text{eff}},
\end{equation}
are zero for all $k=1,...,N_c$. Consequently, $\tilde{o}^{\text{eff}}$ can be
considered an unbiased estimate of $o^{\text{eff}}$. Furthermore,
under the assumption that
\begin{equation}
\text{var}(\epsilon_O(k))=\sigma^2
\end{equation}
for all $k=1,\ldots,N_c$, and
\begin{equation}
\text{cov}(\epsilon_O(k),\epsilon_O(k'))=0
\end{equation}
for all $\{k,k'=1,\ldots,N_c| k \neq  k'\}$, one can define the
variance-covariance matrix as $\sigma^2 B^{-1}= \sigma^2
(c^Tc)^{-1}$, for which the unbiased estimate for $\sigma^2$ is given
by
\begin{equation}
\tilde{\sigma}^2 = \frac {1}{N_c-m} \sum_{k=1}^{N_c} \epsilon_O(k)^2.
\end{equation}
Finally, the unbiased estimate for the variance-covariance matrix for
$\tilde{o}^{\text{eff}}$ is given by $B^{-1}\sigma^2$. In what follows
we do not differentiate between notations for variables and their
estimates.
The least-square procedure  described in this section was used to
determine the effective s.p.\ quadrupole moments
  $\{q_{20,\alpha}^{\text{eff}},q_{22,\alpha}^{\text{eff}}, q_{t,\alpha}^{\text{eff}}, \alpha=1,...,m\}$
and angular momentum alignments $\{j_{\alpha}^{\text{eff}}, \alpha=1,...,m\}$.

\subsection{Method of calculations}
\label{sect-meth-calc}

  The CHF calculations were performed using the code {\tt HFODD (v1.75)\/}
\cite{DD.97a,DD.97b} with the interaction SLy4 \cite{Cha98}.
The accuracy of the harmonic oscillator (HO) expansion depends on the
frequencies ($\hbar\omega_x$, $\hbar\omega_y$, and $\hbar\omega_z$) of
the oscillator wave functions and the number $M$ of the HO states included in the
basis. The basis set includes the lowest $M$ states with energies given
by
\begin{equation}
\varepsilon_{n_x,n_y,n_z} =   \hbar \omega_x (n_x + \frac 12)
                        + \hbar \omega_y (n_y + \frac 12)
                        + \hbar \omega_z (n_z + \frac 12).
\end{equation}
An axially symmetric basis
($\omega_x=\omega_y$) with the deformation $q=\omega_x/\omega_z=1.81$,
oscillator frequency $\hbar\omega_0=41A^{-1/3}$ MeV, and value of $M=296$ was
used in all the CHF calculations. This basis provides sufficient numerical
accuracy for the physical observables of interest \cite{Mladen-thesis}.

The CRMF calculations were performed using the computer code developed in
Refs.\ \cite{KR.89,KR.93,AKR.96}. An anisotropic three-dimensional harmonic
oscillator basis with deformation ($\beta_0=0.4,\,\gamma=0^{\circ}$) has been
used in the CRMF calculations. All fermionic states below the energy cutoff
$E^{\mbox{\scriptsize cut-off}}_F \leq 11.5\hbar\omega^F_0$ and all bosonic
states below the energy cutoff $E^{\mbox{\scriptsize cut-off}}_B \leq 16.5\hbar\omega^B_0$
were used in the diagonalization and the matrix inversion. This basis provides
sufficient numerical accuracy. The NL1 parametrization of the RMF Lagrangian \cite{NL1}
has been used in the CRMF calculations. As follows from our previous studies, this
parametrization provides reasonable s.p.\  energies for nuclei around the valley of
$\beta$-stability \cite{ALR.98,A250}.

\subsection{Selection of independent-particle  configurations}
\label{sec-sel-conf}

    In both CHF and CRMF calculations, the set of independent-particle  configurations
in nuclei around $^{131}$Ce was considered. The final sets used in
additivity analysis consisted of 183 and 105 configurations in the
CHF and CRMF variants, respectively. All ambiguous cases, due to
crossings, convergence difficulties, etc.,  were removed from those
sets. Since the CRMF calculations are more time-consuming than the
CHF ones, the CRMF set is smaller. Nonetheless, the adopted CRMF  set is
sufficiently large to provide reliable results. To put things in perspective,
in Ref.\
\cite{SDDN.96}, where the CHF analysis of additivity principle in the
SD bands of the $A\sim 150$ mass region was carried out,
74 calculated SD configurations were considered.

Every calculated product-state configuration was labeled using the standard notation in
terms of  parity-signature blocks $\left[ N_{+,+i}, N_{+,-i}, N_{-,+i}, N_{-,-i}
\right]$, where $N_{\pi,r}$ are the numbers of occupied s.p.\  orbitals
having  parity $\pi$ and signature  $r$. In addition, the s.p.\ states were labeled
by the Nilsson quantum numbers and signature $[{\cal N}n_z\Lambda]\Omega^{r}$
 of the active orbitals at zero frequency. The orbital identification
 is relatively straightforward
when  the s.p.\  levels do not cross (or cross with a small interaction matrix element),
 but it can become ambiguous  when the crossings with
strong mixing occur.  In some cases,
it was necessary to construct  diabatic routhians by removing weak interaction at 
crossing points. Even with these
precautions, a reliable configuration assignment was not always possible;
the exceptional cases  were excluded from the additivity analysis.  Clearly,
the likelihood of the occurrence
of level crossings is reduced when the s.p.\ level density is small, e.g.,
in the vicinity of large shell gaps.

Large deformed energy gaps develop at high rotational velocity for $Z$=58 and $N$=73
(see Figs.\ \ref{131Ce-sp-neut} and \ref{131Ce-sp-prot}). Therefore, the lowest
SD band ($\nu i_{13/2}$ band) in $^{131}$Ce is a natural
choice for the highly deformed core configuration in the $A$$\sim$130 mass region.
The
additivity analysis was performed at a large rotational frequency of $\hbar\omega$=0.65\,MeV.
 This choice was dictated by the fact that (i) at this frequency the
pairing is already considerably quenched, and (ii) no level crossings appear
in the core  configuration around this frequency  (cf. Figs.\
\ref{131Ce-sp-neut} and \ref{131Ce-sp-prot}). Moreover, at this frequency, the lowest
neutron $i_{13/2}$ intruder orbital already appears below the $N$=73 neutron gap
(see Fig.\ \ref{131Ce-sp-neut}). The choice of  an odd-even core,
strongly motivated by its doubly 
closed character at large deformations/spins, does not impact the
additivity scheme, which is insensitive to the selection of the  reference system.

The highly deformed core configuration in $^{131}$Ce  ([18,19,18,18]$_n\otimes$ [14,14,15,15]$_p$)
has the following orbital structure:
\begin{eqnarray}
\left| {\mbox{\bf core}} \right> =
 \left| {\mbox{\bf core}} \right>_\nu & \otimes &
\left| {\mbox{\bf core}} \right>_\pi
\equiv \nonumber \\[4pt]
  \begin{array}{r}
        (\nu (1i_{13/2}) {\mbox{\bf 6}}_1^{-i} ) \\[-1pt]
        (\nu (1h_{11/2}) [523]7/2^{\pm i})^2 \\[-1pt]
        (\nu (1s_{1/2}) [411]1/2^{\pm i})^2 \\[-1pt]
        (\nu (1g_{7/2}) [413]5/2^{\pm i})^2 \\[-1pt]
        (\nu (2f_{7/2}) [541]1/2^{\pm i})^2 \\[-1pt]
        (\nu (2d_{5/2}) [411]3/2^{\pm i})^2 \\[-1pt]
        (\nu (1h_{11/2}) [532]5/2^{\pm i})^2 \\[-1pt]
        (\nu (1g_{9/2}) [404]9/2^{\pm i})^2 \\[1pt]
        (~\cdots~) \left|{\mbox{\bf 0}}\right>_{\nu}
  \end{array}
&\otimes&
  \begin{array}{r}
        (\pi (1g_{9/2}) [404]9/2^{\pm i})^2 \\
        (\pi (1h_{11/2}) [541]3/2^{\pm i})^2 \\
        (\pi (1g_{7/2}) [420]1/2^{\pm i})^2 \\
        (\pi (2d_{5/2}) [422]3/2^{\pm i})^2 \\
        (\pi (1h_{11/2}) [550]1/2^{\pm i})^2 \\
        (\pi (2p_{1/2})[301]1/2^{\pm i})^2 \\[1pt]
        (~\cdots~) \left|{\mbox{\bf 0}}\right>_{\pi},
  \end{array} \nonumber \label{eqncore}
\end{eqnarray}
where dots denote the deeply bound states and
$\left|{\mbox{\bf 0}}\right>_{\nu}$ and $\left|{\mbox{\bf 0}}\right>_{\pi}$
are the neutron and proton vacua, respectively.
The spherical subshells from which the deformed s.p.\  orbitals emerge
(cf.\ Fig.\ \ref{Nils-diag-Ba128}) are indicated in the front of the Nilsson labels.

  The Nilsson orbital content of an excited configuration is given in
terms of particle and hole excitations with respect to the core configuration
through the action of particle/hole operators with quantum labels corresponding
to the occupied or emptied Nilsson orbitals. The character of the orbital
(particle or hole) is defined by the position of the orbital with respect to
the Fermi level of the core configuration. It is clear from Fig.\ \ref{131Ce-sp-neut}
that the neutron states  $\nu [523]7/2^{\pm i}$, $\nu [411]1/2^{\pm i}$, $\nu [413]5/2^{\pm i}$,
$\nu [541]1/2^{\pm i}$, $\nu [532]5/2^{\pm i}$ and $\nu 6^{-i}_1$
have hole character, while $\nu 6^{+i}_2$, $\nu 6^{-i}_3$, $\nu [530]1/2^{\pm i}$,
$\nu [402]5/2^{\pm i}$, $\nu [532]3/2^{\pm i}$, and $\nu [514]9/2^{\pm i}$ have
particle character. In a similar way, the proton  orbitals $\pi [541]3/2^{\pm i}$,
$\pi [422]3/2^{\pm i}$, $\pi [301]1/2^{\pm i}$, and $\pi [420]1/2^{\pm i}$
and $\pi [404]9/2^{\pm i}$ can be viewed as holes, while
$\pi [532]5/2^{\pm i}$, $\pi [411]3/2^{\pm i}$, $\pi [541]1/2^{\pm i}$, and 
$\pi [413]5/2^{\pm}$ have  particle character (see Fig.\ \ref{131Ce-sp-prot}).

\section{Results of the additivity analysis}
\label{res-add}

\subsection{Effective charge quadrupole moments {\boldmath$q^{\text{eff}}_{20,\alpha}$\unboldmath}}

Table \ref{tbl_q20eff} contains the values of CHF and CRMF
effective s.p.\ charge
quadrupole moments $q^{\text{eff}}_{20,\alpha}$ for a number of s.p.
orbitals  in the vicinity of the deformed
shell gaps at $Z$=58 and $N$=73
(see Figs.\ \ref{131Ce-sp-neut} and \ref{131Ce-sp-prot}). There is an
overall excellent agreement between  $q_{20,\alpha}^{\text{eff}}$
values for the two mean-field approaches employed. In the
majority of cases, the uncertainties are small enough to allow determination
of effective moments to two significant digits.

\newcommand{\ph}{\phantom{$-$}}
\begin{table*}[htp]
\caption{Effective s.p.\  charge quadrupole moments
$q^{\text{eff}}_{20,\alpha}$ (in eb) for the s.p.\  orbitals active
in the $A$$\sim$130 mass region of high- and superdeformation.
Calculations  were carried out
  with CHF+SLy4 and CRMF+NL1 approaches. The bare quadrupole
  moments  $q^{\text{bare}}_{20,\alpha}$ are also shown for CHF+SLy4.
  Theoretical errors resulting from the multivariate least-square fit
  are indicated.  The results of previous calculations
  \cite{SDDN.96} pertaining to  the $A$$\sim$150 mass region are  displayed
  for comparison.}
\label{tbl_q20eff}  
\begin{center}
\begin{tabular}{c|ccc|ccr@{$\pm$}l |r@{$\pm$}l}
\hline \\[-11pt] \hline\\[-13pt] 
     State & & {\small CHF+SkP} \qquad & \quad {\small CHF+SkM*} &
         \multicolumn{4}{|c}{CHF+SLy4}  &   \multicolumn{2}{|c}{CRMF+NL1} \\[3pt
]
        [${\cal N}n_z\Lambda$]$\Omega^r$ & {\footnotesize $p/h$} & $q_{20,\alpha}^{\text{eff}} $ & $q_{20,\alpha}^{\text{eff}}$ &
          {\footnotesize $p/h$} & $q^{\text{bare}}_{20,\alpha}$ &
          \multicolumn{2}{c}{$q_{20,\alpha}^{\text{eff}}$} &
          \multicolumn{2}{|c}{$q_{20,\alpha}^{\text{eff}}$} \\[2pt]
\cline{1-1} \cline{2-4} \cline{5-10} 
     $\nu$ [402]$\frac{5}{2}^{+i}$ &$p$& --0.44 & --0.38 &$p$& 0.0 & --0.35 & 0.01 & --0.26 & 0.01 \\[-1pt]
     $\nu$ [402]$\frac{5}{2}^{-i}$ &$p$& --0.44 & --0.38 &$p$& 0.0 & --0.34 & 0.02 & --0.26 & 0.02 \\[-1pt]
     $\nu$ [411]$\frac{1}{2}^{+i}$ &$h$&        & --0.18 &$h$&   \,\,\,0.0 & --0.15 & 0.02 & --0.11 & 0.02 \\[-1pt]
     $\nu$ [411]$\frac{1}{2}^{-i}$ &$h$&        & --0.15 &$h$&   \,\,\,0.0 & --0.12 & 0.01 & --0.06 & 0.02 \\[-1pt]
     $\nu$ [411]$\frac{3}{2}^{+i}$ &$h$&        & &$h$&   \,\,\,0.0 & --0.15 & 0.04 & --0.13 & 0.03 \\[-1pt]
     $\nu$ [411]$\frac{3}{2}^{-i}$ &$h$&        & &$h$&   \,\,\,0.0 & --0.11 & 0.05 & --0.12 & 0.03 \\[-1pt]
     $\nu$ [413]$\frac{5}{2}^{+i}$ &$h$& --0.16 & &$h$&   \,\,\,0.0 & --0.13 & 0.02 & --0.13 & 0.03 \\[-1pt]
     $\nu$ [413]$\frac{5}{2}^{-i}$ &$h$& --0.13 & &$h$&   \,\,\,0.0 & --0.12 & 0.03 & --0.11 & 0.02 \\[4pt]
     $\nu$  [523]$\frac{7}{2} ^{+i}$ &&& &$h$& \,\,\,0.0 &       0.03 & 0.01 & 0.05 & 0.01 \\[-1pt]
     $\nu$  [523]$\frac{7}{2} ^{-i}$ &&& &$h$& \,\,\,0.0 &       0.04 & 0.01 & 0.01 & 0.02 \\[-1pt]
     $\nu$  [530]$\frac{1}{2} ^{+i}$ &&& &$p$& \,\,\,0.0 &       0.22 & 0.01 & 0.17 & 0.01 \\[-1pt]
     $\nu$  [530]$\frac{1}{2} ^{-i}$ &&& &$p$& \,\,\,0.0 &       0.17 & 0.01 & 0.19 & 0.01 \\[-1pt]
     $\nu$  [532]$\frac{3}{2} ^{+i}$ &&& &$p$& \,\,\,0.0 &       0.21 & 0.03 & \multicolumn{2}{c}{---}  \\[-1pt]
     $\nu$  [532]$\frac{3}{2} ^{-i}$ &&& &$p$& \,\,\,0.0 &       0.17 & 0.03 & \multicolumn{2}{c}{---}  \\[-1pt]
     $\nu$  [532]$\frac{5}{2} ^{+i}$ &&& &$h$& \,\,\,0.0 &       0.19 & 0.03 & 0.17 & 0.03 \\[-1pt]
     $\nu$  [532]$\frac{5}{2} ^{-i}$ &&& &$h$& \,\,\,0.0 &       0.24 & 0.03 & 0.38 & 0.03 \\[-1pt]
     $\nu$  [541]$\frac{1}{2} ^{+i}$ &&& &$h$& \,\,\,0.0 &       0.35 & 0.03 & 0.35 & 0.02 \\[-1pt]
     $\nu$  [541]$\frac{1}{2} ^{-i}$ &&& &$h$& \,\,\,0.0 &       0.37 & 0.03 & 0.33 & 0.03 \\[4pt]
     $\nu$  {\bf 6}$_1^{-i}$         &&& &$h$& \,\,\,0.0 &       0.38 & 0.01 & 0.40 & 0.01 \\[-1pt]
     $\nu$  {\bf 6}$_2^{+i}$         &&& &$p$& \,\,\,0.0 &       0.36 & 0.01 & 0.36 & 0.01 \\[-1pt]
     $\nu$  {\bf 6}$_3^{-i}$  &$h$& \,\,\,0.43 & \,\,\,0.30  &$p$& \,\,\,0.0 & 0.35 & 0.05 & \multicolumn{2}{c}{---} \\[2pt]
\hline
 $\pi$ [301]$\frac{1}{2}^{+i}$ &$h$&--0.15 & --0.13 &$h$& --0.08 & 0.51 & 0.05 & \multicolumn{2}{c}{---} \\[4pt]
 $\pi$ [404]$\frac{9}{2}^{+i}$ &$p$&--0.30 & --0.28 &$p$& --0.13 & --0.32 & 0.01 &   --0.37 & 0.01 \\[-1pt]
 $\pi$ [404]$\frac{9}{2}^{-i}$ &$p$&--0.30 & --0.28 &$p$& --0.13 & --0.32 & 0.01 &   --0.37 & 0.01 \\[-1pt]
 $\pi$ [411]$\frac{3}{2}^{+i}$ &$p$& \,\,\,0.11 & \,\,\,0.10 &$p$& \,\,\,0.06 & --0.05 & 0.02 & \multicolumn{2}{c}{---}  \\[-1pt]
 $\pi$ [411]$\frac{3}{2}^{-i}$ &$p$& \,\,\,0.11 & \,\,\,0.10 &$p$& \,\,\,0.06 & 0.00 & 0.01 & \multicolumn{2}{c}{---}  \\[-1pt]
 $\pi$ [413]$\frac{5}{2}^{-i}$ &&& &$p$& \,\,\,0.06 &     0.28 & 0.05 &   \multicolumn{2}{c}{---}  \\[-1pt]
 $\pi$ [422]$\frac{3}{2}^{+i}$ &&& &$h$& \,\,\,0.20 &       0.33 & 0.02 &       0.33
 & 0.03 \\[-1pt]
 $\pi$ [422]$\frac{3}{2}^{-i}$ &&& &$h$& \,\,\,0.22 &       0.34 & 0.02 &       0.28  & 0.02 \\[4pt]
 $\pi$ [532]$\frac{5}{2} ^{+i}$ &&& &$p$& \,\,\,0.28 &       0.43 & 0.01 &       0.41 & 0.02 \\[-1pt]
 $\pi$ [532]$\frac{5}{2} ^{-i}$ &&& &$p$& \,\,\,0.36 &       0.56 & 0.03 &       0.54 & 0.03 \\[-1pt]
 $\pi$ [541]$\frac{1}{2} ^{-i}$ &&& &$p$& \,\,\,0.40 &       0.58 & 0.02 &   \multicolumn{2}{c}{---} \\[-1pt]
 $\pi$ [541]$\frac{3}{2} ^{+i}$ &&& &$h$& \,\,\,0.34 &       0.50 & 0.01 &       0.48 & 0.01 \\[-1pt]
 $\pi$ [541]$\frac{3}{2} ^{-i}$ &&& &$h$& \,\,\,0.39 &       0.57 & 0.01 &       0.50 & 0.01 \\[-1pt]
 $\pi$ [550]$\frac{1}{2} ^{-i}$ &&& &$h$& \,\,\,0.30 &       0.49 & 0.05 &       0.47 & 0.04 \\[2pt]
 \hline \hline
\end{tabular}
\end{center} 
\end{table*}

The two lowest neutron intruder orbitals
{\bf 6}$_1^{-i}$ and {\bf 6}$_2^{+i}$
 show significant
signature splitting, and their effective charge quadrupole moment values
differ by more than 5\%. The extracted values confirm the general
expectations
for the polarization effects exerted by the intruder and extruder states
\cite{FM83,CFL83}. The
lowest neutron ${\cal N}$=6 orbitals, {\bf 6}$_1^{-i}$ and {\bf 6}$_2^{+i}$,
have $q_{20,\alpha}^{\text{eff}}\simeq$ 0.37\,eb, which indicates that
their occupation  drives the nucleus towards larger prolate deformation.
The third intruder orbital, {\bf 6}$_3^{-i}$, although calculated with
relatively poor statistics, confirms this trend.

The proton $\pi [404]9/2^{\pm i}$ extruder high-$\Omega$
orbitals are oblate-driving; they have large negative
values of $q_{20,\alpha}^{\text{eff}}$.  Emptying them
polarizes the nucleus  towards more prolate-deformed shapes.
Interestingly, their $q_{20,\alpha}^{\text{eff}}$ values of around
$-0.31$\,eb are close in magnitude to those of the ${\cal N}$=6 neutron
intruders, in line with the findings of Ref.\ \cite{AR.96} that the holes
in the proton $g_{9/2}$ orbitals are as important as the particles in
the neutron $i_{13/2}$ orbitals in stabilizing the shape at large
deformation. Due to their high-$\Omega$ content, the signature splitting of
 $\pi [404]9/2^{\pm i}$ routhians is extremely small
  and their $q_{20,\alpha}^{\text{eff}}$  values are
practically indistinguishable within error bars.

Our study indicates that proton $h_{11/2}$ states, such as $\pi [541]3/2^{\pm i}$
and $\pi [532]5/2^{\pm i}$ active below and above the $Z$=58  shell gap,
respectively, play a significant role in the existence of this island of high
deformation. Indeed, Table\ \ref{tbl_q20eff} attributes them to  effective charge
quadrupole moments in excess of 0.45\,eb - very significant values compared
with other states listed.

The downsloping orbital $\pi [541]1/2^{-i}$, originating from mixed $\pi
f_{7/2}\oplus h_{9/2}$ subshells, carries a large effective charge
quadrupole moment of more than 0.5\,eb. Although one could expect it to
play a role in the formation of large prolate deformation, this state
appears too high in energy (above the $Z$=58 shell gap) and would
therefore always stay unoccupied in most of the configurations of
interest \cite{A130-exp1,A130-exp2,Ce132-131}. On the contrary, the
strongly prolate-driving $\pi [550]1/2^{-i}$ orbital carrying
$q_{20,\alpha}^{\text{eff}}$$\approx$0.47\,eb, is always occupied in the
bands of interest.

Table \ref{tbl_q20eff} compares the values of
$q_{20,\alpha}^{\text{eff}}$ obtained in the present study with those
from the additivity analysis of the SD bands in the $A$$\sim$150 region
\cite{SDDN.96} based on the Skyrme SkP and SkM* energy density
functionals. Note that some of the states, which are of particle
character  in the $A$$\sim$130 region, appear as hole states in the
heavier region. For these states, conforming to our definitions of
coefficients $c_{\alpha}$ (Sec.\ \ref{add-prin-math}), we inverted signs
of values shown in Table~1 of Ref.\ \cite{SDDN.96}. With few exceptions,
$q_{20,\alpha}^{\text{eff}}$ values are similar in both studies: only
for the $\pi [301]1/2^{+i}$ orbital does the difference between $A\sim 130$
and $A\sim 150$ results  exceed 0.1\,eb. This result strongly suggests
that the polarization effects caused by occupying/emptying  specific
orbitals are mainly due to the general geometric properties of s.p.
orbitals and weakly depend on the actual  parametrization of the Skyrme
energy density functional; minor differences are likely related to
interactions between close-lying s.p.\ states. These observations give
strong reasons for combining the two regions into one, and interpreting
the entire area of highly and SD  rotational states in the
mass range $A\sim 128-160$ within the united theoretical framework.

The results for $q_{20,\alpha}^{\text{eff}}$
obtained in  CHF+SLy4
and CRMF+NL1 models  are  indeed very similar (see Table \ref{tbl_q20eff}). Only for
the $\nu
[402]5/2^{\pm i}$ and $\nu [532]5/2^{-i}$ orbitals, do the differences
between  $q_{20,\alpha}^{\text{eff}}$ values come close to 0.1\,eb.

Table \ref{tbl_q20eff} compares the bare and effective s.p.\  charge
quadrupole moments obtained in  CHF+SLy4. In the majority of cases,
these quantities differ drastically, underlying the importance of
shape polarization effects.   Large differences between  bare and
effective s.p.\  quadrupole moments have also been found in the CHF+SkP
and CHF+SkM* calculations in the $A$$\sim$150 region of superdeformation
\cite{SDDN.96}.

\subsection{Effective  quadrupole moments
{\boldmath$q_{t,\alpha}^{\text{eff}}$\unboldmath} and
{\boldmath$q_{22,\alpha}^{\text{eff}}$\unboldmath}}
\label{Sect-q22-qt}

\begin{figure*}
\centering
\includegraphics[width=13cm]{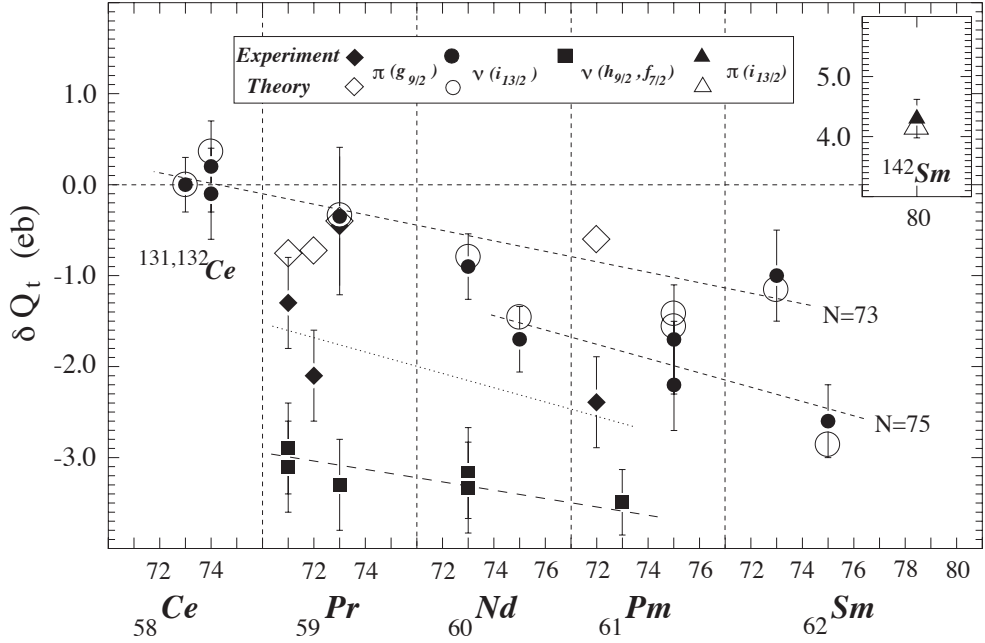}
\caption{Experimental (closed symbols with error bars) and calculated (CRMF+NL1,
open symbols) differential transition quadrupole moments for
highly deformed bands in Ce, Pr, Nd, Pm, and Sm
isotopes. The experimental data were taken from Refs.\
\cite{A130-exp1,A130-exp2} and references quoted therein. 
The values of $\delta Q_t$ for SD band in $^{132}$Sm are shown in the inset.
Dashed lines are drawn
 to guide the eye.}
\label{diff-qt}
\end{figure*}

Table\ \ref{tbl_q22qteff} displays the calculated
effective s.p.\  transition quadrupole moments
$q_{t,\alpha}^{\text{eff}}$, cf.\ definitions
(\ref{Qteff}--\ref{qteff}). Based on the additivity principle, these
values can be used to predict the total charge transition moments
$Q_{t}(k)$ in highly deformed and SD bands  of $A\sim130$ nuclei:
\begin{equation}
\label{Qt-estimated}
Q_{t}(k)=Q_{t}^{\text{core}} + \sum_{\alpha}c_{\alpha}(k) q_{t,\alpha}^{\text{eff}} ,
\end{equation}
where the calculated CHF+SLy4 value for the core configuration
in $^{131}$Cs is
\begin{equation}
Q_{t}^{\text{core}} = 7.64\,\mbox{eb} .
\end{equation}
Since the total calculated values are  less precise than the
relative ones which define the effective s.p.\  transition
quadrupole moments $q_{t,\alpha}^{\text{eff}}$, one may alternatively use
in Eq.\ (\ref{Qt-estimated}) the measured value \cite{Cla96},
\begin{equation}
Q_{t}^{\text{core,exp}} = 7.4(3)\,\mbox{eb}.
\end{equation}
Theoretical estimates of the total charge transition moments
$Q_{t}(k)$ allow  for predictions of  $B(E2)$ values
\begin{equation}
B(E2)(I\rightarrow{I-2},k)= \frac{5}{16\pi} e^2 \langle I0\,20|I-2\,0\rangle Q^2_{t}(k),
\end{equation}
and lifetimes \cite{NR.96}.

In CHF+SLy4, the uncertainties of
$q_{22,\alpha}^{\text{eff}}$ appear to be larger than those for
$q_{20,\alpha}^{\text{eff}}$. In CRMF+NL1, however, those uncertainties
are
similar. This can be traced back to the different  $\gamma$-softness of
 potential energy
surfaces in CHF+SLy4 and CRMF+NL1
 (see Ref.\ \cite{AR.96} and references quoted therein for the results obtained in
different approaches); current analysis revealing large uncertainties
for $q_{22,\alpha}^{\text{eff}}$ suggests that the potential energy
surfaces are softer (and, thus less localized) in the CHF+SLy4 approach.

Although the  values of $q_{22,\alpha}^{\text{eff}}$ are generally much
smaller than $q_{20,\alpha}^{\text{eff}}$,
large uncertainties in the determination of certain moments
$q_{22,\alpha}^{\text{eff}}$ (especially, for  $\nu [411]3/2^{\pm
i}$, $\nu [532]5/2^{\pm i}$, $\nu 6_3^{-i}$, $\pi [301]1/2^{+i}$, and
$\pi [550]1/2^{-i}$ orbitals, for which the errors
exceed 0.1\,eb in the CHF+SLy4 approach) can lead to the deterioration
of predicted $q_{t,\alpha}^{\text{eff}}$. On the other hand, in
many cases the uncertainties in $q_{22,\alpha}^{\text{eff}}$ are smaller
than the experimental error bars; hence,  they are less relevant when
comparison with experiment is carried out.
Currently available experimental data on relative
transition quadrupole moments agree reasonably well with the CHF+SLy4
results \cite{A130-exp1,A130-exp2}.

Table \ref{tbl_q22qteff}  compares  $q_{t,\alpha}^{\text{eff}}$ values
obtained in  CHF+SLy4 and CRMF+NL1 models. The results for proton
orbitals are similar in both approaches: the differences between
respective  $q_{t,\alpha}^{\text{eff}}$ values do not exceed 0.1\,eb.
Larger differences are seen for the neutrons: for about 50 percent  of
calculated orbitals ($\nu [402]5/2^{\pm i}$, $\nu [411]1/2^{\pm i}$,
$\nu [411]3/2^{+i}$, $\nu [413]5/2^{-i}$, $\nu [530]1/2^{+i}$, and $\nu
[532]5/2^{+i}$), the difference between $q_{t,\alpha}^{\text{eff}}$
values  in CHF+SLy4 and CRMF+NL1 exceeds 0.1\,eb. Interaction (mixing)
between those close-lying states (see Fig.\ \ref{131Ce-sp-neut}),
predicted differently in the two approaches, is the most likely reason
for the deviations seen.

The results of CHF+SLy4  were compared with experimental transition
moments  in Refs.\ \cite{A130-exp1,A130-exp2}. Here, we show in Fig.\
\ref{diff-qt} a comparison between  CRMF+NL1 and experiment for the
relative transition quadrupole moments $\delta Q_t(k)$ in different
highly deformed and SD bands in nuclei with $Z$=57-62 involving
$i_{13/2}$ neutrons and/or $g_{9/2}$ proton holes. The agreement between
experiment and theory is quite remarkable with all the
experimental trends discussed in Refs.\ \cite{A130-exp1,A130-exp2}
well reproduced by calculations. One should note that the CRMF and CHF results are close
to each other. The general pattern of decreasing $Q_t$ with increasing $Z$
and $N$ is consistent with the general expectation that as one adds
particles above a deformed shell gap, the deformation-stabilizing effect
of the gap is diminished. This trend continues until a new ``magic''
deformed number is reached. Such a situation  occurs when going from
$^{132}$Ce towards  $Z$=62 and $N$=80 ($^{142}$Sm), where a large jump
in transition quadrupole moment takes place marking the point at which
it becomes energetically favorable to fill the high-$j$ $\pi i_{13/2}$
and $\nu j_{15/2}$ orbitals responsible for the existence of  the $A\sim
142$ SD  island.

It is gratifying to see that  CRMF+NL1 reproduces the value of $Q_t$ in
$^{142}$Sm based on  the $^{131}$Ce core (see inset in Fig.\ \ref{diff-qt}).
Earlier on, it  was demonstrated in Refs.~\cite{A130-exp1,A130-exp2,142Sm-exp}
that  this  $Q_t$ value can be also  reproduced within  CHF
using   either a $^{131}$Ce or a $^{152}$Dy core.

\begin{table*}[htp]
\caption{Effective s.p.\  charge quadrupole moments
$q_{20,\alpha}^{\text{eff}}$ and $q_{22,\alpha}^{\text{eff}}$ as well
as the transition quadrupole moments $q_{t,\alpha}^{\text{eff}}$ (all in eb)
calculated in  CHF+SLy4 and CRMF+NL1.}
\label{tbl_q22qteff}  
\begin{center}
\begin{tabular}{c@{\hspace{2pt}} |r@{$\pm$}l@{\hspace{3pt}} r@{$\pm$}l@{\hspace{3pt
}} r@{$\pm$}l |r@{$\pm$}l@{\hspace{3pt}}r@{$\pm$}l@{\hspace{3pt
}}r@{$\pm$}l} 
\hline \\[-11pt] \hline\\[-13pt]
     State &  \multicolumn{6}{|c}{CSHF~+~SLy4} &  \multicolumn{6}{|c}{CRMF~+~NL1} \\
[1pt]
     [${\cal N}n_z\Lambda$]$\Omega^r$ &
        \multicolumn{2}{c} {$q_{20,\alpha}^{\text{eff}}$}  &
        \multicolumn{2}{c} {$q_{22,\alpha}^{\text{eff}}$}  &
        \multicolumn{2}{c} {$q_{t ,\alpha}^{\text{eff}}$}  &
        \multicolumn{2}{|c}{$q_{20,\alpha}^{\text{eff}}$}  &
        \multicolumn{2}{c} {$q_{22,\alpha}^{\text{eff}}$}  &
        \multicolumn{2}{c} {$q_{t ,\alpha}^{\text{eff}}$}
\\[3pt] \hline
$\nu$[402]$\frac{5}{2} ^{+i}$ &
        --0.35 & 0.01 &   0.14 & 0.06  & --0.04 & 0.04  & 
        --0.26 & 0.02 & --0.02& 0.01  & --0.25& 0.02   \\[-0.5pt] 
$\nu$[402]$\frac{5}{2} ^{-i}$ &
        --0.34 & 0.02 &   0.08 & 0.08  & --0.38 & 0.05  & 
        --0.26 & 0.02 & --0.07& 0.02  & --0.22& 0.03  \\[-0.5pt] 
$\nu$[411]$\frac{1}{2} ^{+i}$ &
        --0.15 & 0.02 & --0.24 & 0.10  & --0.01 & 0.06  & 
        --0.11 & 0.02 &   0.09 & 0.02  & --0.16 & 0.02  \\[-0.5pt] 
$\nu$[411]$\frac{1}{2} ^{-i}$ &
        --0.12 & 0.01 &   0.06 & 0.06  & --0.16 & 0.04  & 
        --0.06 & 0.02 & --0.17 & 0.02  &   0.04 & 0.02  \\[-0.5pt] 
$\nu$[411]$\frac{3}{2} ^{+i}$ &
        --0.15 & 0.04 &   0.20 & 0.20  & --0.26 & 0.12  & 
        --0.13 & 0.03 & --0.02 & 0.03  & --0.11 & 0.03  \\[-0.5pt] 
$\nu$[411]$\frac{3}{2} ^{-i}$ &
        --0.11 & 0.05 & --0.05 & 0.24  & --0.08 & 0.15  & 
        --0.12 & 0.03 &   0.02 & 0.03  & --0.12 & 0.03  \\[-0.5pt] 
$\nu$[413]$\frac{5}{2} ^{+i}$ &
        --0.13 & 0.02 & --0.05 & 0.10  & --0.10 & 0.06  & 
        --0.13 & 0.03 & --0.04 & 0.03  & --0.10 & 0.03  \\[-0.5pt] 
$\nu$[413]$\frac{5}{2} ^{-i}$ &
        --0.12 & 0.03 & --0.12 & 0.13  & --0.05 & 0.08  & 
        --0.11 & 0.02 &   0.15 & 0.03  & --0.20 & 0.03  \\[3pt] 
$\nu$[523]$\frac{7}{2} ^{+i}$ &
          0.03 & 0.01 & --0.00 & 0.05  &   0.03 & 0.03  & 
          0.05 & 0.01 &   0.00& 0.01  &   0.04 & 0.01  \\[-0.5pt] 
$\nu$[523]$\frac{7}{2} ^{-i}$ &
          0.04 & 0.01 & --0.01 & 0.05  &   0.05 & 0.03  & 
          0.01 & 0.02 & --0.00 & 0.02  &   0.01 & 0.02  \\[-0.5pt] 
$\nu$[530]$\frac{1}{2} ^{+i}$ &
          0.22 & 0.01 & --0.21 & 0.05  &   0.34 & 0.03  & 
          0.17 & 0.01 & --0.09 & 0.01  &   0.22 & 0.01  \\[-0.5pt] 
$\nu$[530]$\frac{1}{2} ^{-i}$ &
          0.17 & 0.01 & --0.01 & 0.05  &   0.18 & 0.03  & 
          0.19 & 0.01 &   0.10 & 0.01  &   0.13 & 0.01  \\[-0.5pt] 
$\nu$[532]$\frac{3}{2} ^{+i}$ &
          0.21 & 0.03 &   0.21 & 0.13  &   0.09 & 0.08  & 
           \multicolumn{2}{c}{---}   &   \multicolumn{2}{c}{---}    &  \multicolumn{2}{c}{---} \\[-0.5pt]
$\nu$[532]$\frac{3}{2} ^{-i}$ &
          0.17 & 0.03 &   0.03 & 0.13  &   0.15 & 0.08  & 
           \multicolumn{2}{c}{---}   &  \multicolumn{2}{c}{---}    &  \multicolumn{2}{c}{---} \\[-0.5pt]
$\nu$[532]$\frac{5}{2} ^{+i}$ &
          0.19 & 0.03 & --0.08 & 0.20  &   0.24 & 0.12  & 
          0.17 & 0.03 & --0.02 & 0.03  &   0.18 & 0.03  \\[-0.5pt] 
$\nu$[532]$\frac{5}{2} ^{-i}$ &
          0.24 & 0.03 & --0.01 & 0.20  &   0.25 & 0.12  & 
          0.38 & 0.03 &   0.00 & 0.03  &   0.38 & 0.03  \\[-0.5pt] 
$\nu$[541]$\frac{1}{2} ^{+i}$ &
          0.35 & 0.03 & --0.04 & 0.13  &   0.38 & 0.08  & 
          0.35 & 0.02 & --0.00 & 0.02  &   0.35 & 0.03  \\[-0.5pt] 
$\nu$[541]$\frac{1}{2} ^{-i}$ &
          0.37 & 0.03 &   0.01 & 0.14  &   0.36 & 0.08  & 
          0.33 & 0.03 &   0.04 & 0.03  &   0.30 & 0.03  \\[3pt] 
$\nu$  {\bf 6}$_{1} ^{-i}$ &
          0.38 & 0.01 &   0.21 & 0.03  &   0.26 & 0.02  & 
          0.40 & 0.01 &   0.12 & 0.01  &   0.33 & 0.01  \\[-0.5pt] 
$\nu$  {\bf 6}$_{2} ^{+i}$ &
          0.36 & 0.01 & --0.01 & 0.04  &   0.37 & 0.03  & 
          0.36 & 0.01 & --0.01 & 0.01  &   0.37 & 0.01  \\[-0.5pt] 
$\nu$  {\bf 6}$_{3} ^{-i}$ &
          0.35 & 0.05 & --0.06 & 0.22  &   0.38 & 0.13  & 
           \multicolumn{2}{c}{---}   &   \multicolumn{2}{c}{---}    &   \multicolumn{2}{c}{---} \\[2pt]
 \hline
$\pi$[301]$\frac{1}{2} ^{+i}$ &
          0.51 & 0.05 & --0.10 & 0.24  &   0.57 & 0.14  & 
           \multicolumn{2}{c}{---}   &  \multicolumn{2}{c}{---}    &   \multicolumn{2}{c}{---} \\[3pt]
$\pi$[404]$\frac{9}{2} ^{+i}$ &
        --0.32 & 0.01 &   0.10 & 0.04  & --0.38 & 0.02  & 
        --0.37 & 0.01 &   0.02 & 0.01  & --0.38 & 0.01  \\[-0.5pt] 
$\pi$[404]$\frac{9}{2} ^{-i}$ &
        --0.32 & 0.01 &   0.09 & 0.04  & --0.37 & 0.02  & 
        --0.37 & 0.01 &   0.02 & 0.01  & --0.38 & 0.01  \\[-0.5pt] 
$\pi$[411]$\frac{3}{2} ^{+i}$ &
        --0.05 & 0.02 &   0.10 & 0.07  & --0.10 & 0.05  & 
           \multicolumn{2}{c}{---}   &  \multicolumn{2}{c}{---}    &  \multicolumn{2}{c}{---} \\[-0.5pt]
$\pi$[411]$\frac{3}{2} ^{-i}$ &
          0.00 & 0.01 & --0.22 & 0.07  &   0.12 & 0.04  & 
          \multicolumn{2}{c}{---}  &   \multicolumn{2}{c}{---}   &   \multicolumn{2}{c}{---} \\[-0.5pt]
$\pi$[422]$\frac{3}{2} ^{+i}$ &
          0.33 & 0.02 & --0.27 & 0.10  &   0.48 & 0.06  & 
          0.33 & 0.03 & --0.13 & 0.02  &   0.40 & 0.03  \\[-0.5pt] 
$\pi$[422]$\frac{3}{2} ^{-i}$ &
          0.34 & 0.02 &   0.14 & 0.10  &   0.25 & 0.06  & 
          0.28 & 0.02 &   0.16 & 0.02  &   0.19 & 0.02  \\[3pt] 
$\pi$[532]$\frac{5}{2} ^{+i}$ &
          0.43 & 0.01 & --0.05 & 0.05  & --0.46 & 0.03  & 
          0.41 & 0.02 & --0.04 & 0.01  &   0.43 & 0.02  \\[-0.5pt] 
$\pi$[532]$\frac{5}{2} ^{-i}$ &
          0.56 & 0.03 & --0.07 & 0.09  &   0.60 & 0.05  & 
          0.54 & 0.03 &   0.05 & 0.03  &   0.51 & 0.04  \\[-0.5pt] 
$\pi$[541]$\frac{1}{2} ^{-i}$ &
          0.58 & 0.02 & --0.01 & 0.10  &   0.59 & 0.06  & 
          \multicolumn{2}{c}{---}    & \multicolumn{2}{c}{---}   &  \multicolumn{2}{c}{---} \\[-0.5pt]
$\pi$[541]$\frac{3}{2} ^{+i}$ &
          0.50 & 0.01 & --0.05 & 0.06  &   0.52 & 0.04  & 
          0.48 & 0.01 & --0.10 & 0.01  &   0.54 & 0.01  \\[-0.5pt] 
$\pi$[541]$\frac{3}{2} ^{-i}$ &
          0.57 & 0.01 & --0.12 & 0.04  &   0.63 & 0.03  & 
          0.50 & 0.01 & --0.10 & 0.01  &   0.56 & 0.01  \\[-0.5pt] 
$\pi$[550]$\frac{1}{2} ^{-i}$ &
          0.49 & 0.05 & --0.06 & 0.22 & 0.52 & 0.14  &
          0.47 & 0.04 &--0.02 & 0.04 & 0.48 & 0.04  \\[2pt] 
  \hline \hline
\end{tabular}
\end{center}
\end{table*}

\begin{table}[htp]
\caption{Effective s.p.\  angular momentum alignments  $j^{\text{eff}}_{\alpha}$
(in $\hbar$)  of
the active orbitals calculated in  CHF+SLy4 and CRMF+NL1.
In the second column, the bare s.p.\  angular momenta $j^{\text{bare}}_{\alpha}$,
calculated with  CHF+SLy4  are also shown.
}
\label{tbl_Ieff}  
\begin{center}
\begin{tabular}{c|cr@{$\pm$}l|r@{$\pm$}l} 
\hline \\[-11pt] \hline\\[-13pt]
     State &   \multicolumn{3}{|c}{CHF+SLy4}  & \multicolumn{2}{|c}{CRMF+NL1} \\
[-1pt] 
       [${\cal N}n_z\Lambda$]$\Omega^r$   &
        $j^{\text{bare}}_{\alpha}$ &
      \multicolumn{2}{c} {$j^{\text{eff}}_{\alpha}$} &
      \multicolumn{2}{|c}{$j^{\text{eff}}_{\alpha}$} \\[2pt]
  \hline
       $\nu$  [402]$\frac{5}{2} ^{+i}$ &  $-$0.528 &  0.58 & 0.14 &  0.47 & 0.15\\[-1pt]
       $\nu$  [402]$\frac{5}{2} ^{-i}$ &  $-$0.493 &  0.51 & 0.20 &  0.38 & 0.26\\[-1pt]
       $\nu$  [411]$\frac{1}{2} ^{+i}$ &    \ph0.411 &  0.67 & 0.24 &  0.64 & 0.17\\[-1pt]
       $\nu$  [411]$\frac{1}{2} ^{-i}$ &    \ph0.380 &  0.40 & 0.15 &  0.09 & 0.16\\[-1pt]
       $\nu$  [411]$\frac{3}{2} ^{+i}$ &  $-$0.092 &  1.72 & 0.46 &  1.35 & 0.29\\[-1pt]
       $\nu$  [411]$\frac{3}{2} ^{-i}$ &    \ph0.077 &  0.56 & 0.57 &  1.08 & 0.29\\[-1pt]
       $\nu$  [413]$\frac{5}{2} ^{+i}$ &  $-$0.316 & $-$0.10 & 0.23 &  0.44 & 0.27\\[-1pt]
       $\nu$  [413]$\frac{5}{2} ^{-i}$ &  $-$0.428 &  0.12 & 0.30 &  0.14 & 0.26\\[4pt]
       $\nu$  [523]$\frac{7}{2} ^{+i}$ &  $-$0.908 & $-$1.10 & 0.10 &$-$1.24 & 0.12\\[-1pt]
       $\nu$  [523]$\frac{7}{2} ^{-i}$ &  $-$0.974 & $-$1.19 & 0.12 &$-$0.92 & 0.18\\[-1pt]
       $\nu$  [530]$\frac{1}{2} ^{+i}$ &    \ph1.548 &  1.19 & 0.11 &  1.86 & 0.09\\[-1pt]
       $\nu$  [530]$\frac{1}{2} ^{-i}$ &    \ph0.564 &  0.88 & 0.11 &  0.93 & 0.10\\[-1pt]
       $\nu$  [532]$\frac{3}{2} ^{+i}$ &    \ph0.171 & $-$0.34 & 0.30 &   \multicolumn{2}{c}{---}   \\[-1pt]
       $\nu$  [532]$\frac{3}{2} ^{-i}$ &    \ph0.835 &  0.44 & 0.31 &   \multicolumn{2}{c}{---}   \\[-1pt]
       $\nu$  [532]$\frac{5}{2} ^{+i}$ &  $-$0.331 & $-$0.89 & 0.46 &$-$0.95 & 0.29\\[-1pt]
       $\nu$  [532]$\frac{5}{2} ^{-i}$ &    \ph0.417 & $-$1.06 & 0.46 &$-$1.29 & 0.30\\[-1pt]
       $\nu$  [541]$\frac{1}{2} ^{+i}$ &    \ph1.793 &  0.92 & 0.31 &  0.95 & 0.25\\[-1pt]
       $\nu$  [541]$\frac{1}{2} ^{-i}$ &    \ph0.466 &  0.89 & 0.32 &$-$0.34 & 0.28\\[4pt]
       $\nu$  {\bf 6}$_1 ^{-i}$ &   \ph4.840 &     4.78 & 0.08 &  4.59 & 0.08\\[-1pt]
       $\nu$  {\bf 6}$_2 ^{+i}$ &   \ph4.031 &     3.42 & 0.11 &  3.15 & 0.10\\[-1pt]
       $\nu$  {\bf 6}$_3 ^{-i}$ &   \ph2.662 &     0.77 & 0.50 &   \multicolumn{2}{c}{---}   \\[1pt]
 \hline
       $\pi$  [301]$\frac{1}{2} ^{+i}$ & $-$0.432 &  1.23 & 0.55 &   \multicolumn {2}{c}{---}   \\[-1pt]
       $\pi$  [404]$\frac{9}{2} ^{+i}$ & $-$0.719 &$-$0.00 & 0.09 &  0.09 & 0.09\\[-1pt]
       $\pi$  [404]$\frac{9}{2} ^{-i}$ & $-$0.719 &$-$0.00 & 0.09 &  0.11 & 0.09\\[-1pt]
       $\pi$  [411]$\frac{3}{2} ^{+i}$ & $-$0.249 &  0.81 & 0.18 &   \multicolumn{2}{c}{---}   \\[-1pt]
       $\pi$  [411]$\frac{3}{2} ^{-i}$ & $-$0.062 &  0.65 & 0.16 &   \multicolumn{2}{c}{---}   \\[-1pt]
       $\pi$  [413]$\frac{5}{2} ^{-i}$ & $-$0.539 &$-$1.52 & 0.53 &   \multicolumn{2}{c}{---}   \\[-1pt]
       $\pi$  [422]$\frac{3}{2} ^{+i}$ & $-$0.315 
&$-$0.19 & 0.25 &$-$0.21 & 0.27\\[-1pt]
       $\pi$  [422]$\frac{3}{2} ^{-i}$ &   \ph0.510 
&$-$0.84 & 0.23 &$-$0.38 & 0.24\\[-1pt]
       $\pi$  [532]$\frac{5}{2} ^{+i}$ & $-$0.253 &$-$0.90 & 0.13 &$-$1.11 & 0.16\\[-1pt]
       $\pi$  [532]$\frac{5}{2} ^{-i}$ & $-$0.022 &$-$0.67 & 0.20 &  \multicolumn{2}{c}{---}    \\[-1pt]
       $\pi$  [541]$\frac{1}{2} ^{-i}$ &   \ph0.944 &  1.75 & 0.23 &   \multicolumn{2}{c}{---}   \\[-1pt]
       $\pi$  [541]$\frac{3}{2} ^{+i}$ &   \ph1.743 &  1.57 & 0.13 &  1.18 & 0.11\\[-1pt]
       $\pi$  [541]$\frac{3}{2} ^{-i}$ & $-$0.057 &$-$0.54 & 0.10 &$-$0.48 & 0.11\\[-1pt]
       $\pi$  [550]$\frac{1}{2} ^{-i}$ &   \ph2.819 &  2.99 & 0.52 &  2.86 & 0.40\\[2pt]
\hline \hline
\end{tabular}
\end{center}
\end{table}

\subsection{Effective angular momenta
{\boldmath$j^{\text{eff}}_{\alpha}$\unboldmath} (s.p.\ alignments)}
\label{Sect-jeff}

In this section, we evaluate and interpret the effective s.p.\
contributions to the total angular momentum. Table\ \ref{tbl_Ieff}
displays effective s.p.\  angular momenta $j^{\text{eff}}_{\alpha}$ for
the s.p.\ orbitals of interest. The relative uncertainties in calculated
$j^{\text{eff}}_{\alpha}$ values are on average larger than those for
effective s.p.\  quadrupole moments. This is due to the fact that, on the
mean field level,  polarization effects pertaining to the angular
momentum are more complex than those for  quadrupole moments: they
involve not only shape changes but also the variations of time-odd mean
fields \cite{DD.95,AR.00,VALR.05}. For eight  proton states calculated
in both approaches,  the mean uncertainties are $0.19\hbar$ and
$0.18\hbar$ in  CHF+SLy4 and CRMF+NL1, respectively. The same holds also
for the set of 18 neutron states, where the average uncertainties are
$0.25\hbar$ and $0.20\hbar$ in  CHF+SLy4 and CRMF+NL1, respectively.

Table\ \ref{tbl_Ieff} also compares  CHF+SLy4 expectation values of the
s.p.\  angular momentum $j^{\text{bare}}_{\alpha}=\left<\hat{j}\right>_{\alpha}$ with
their effective counterparts $j^{\text{eff}}_{\alpha}$.
It is seen that these two quantities differ considerably.
As discussed in Ref.~\cite{AR.00},
this  is due to  both  shape polarization and
time-odd mean-field effects. It is also
important to remember that, unlike the cranked Nilsson scheme,
in self-consistent models the expectation value
of the projection of the s.p.\  angular momentum on the rotation
axis $j^{\text{bare}}_{\alpha}$ cannot be extracted from the slope of its
s.p.\  routhian versus rotation frequency \cite{Gal.94}.

Our results indicate that the additivity principle for angular momentum
alignment does not work as precisely as it does for  quadrupole moments.
This conclusion is in line with a similar analysis in the $A\sim 60$
region of superdeformation \cite{D.98,A6080}. A  configuration
assignment based on relative alignments depends on how accurately these
alignments can be predicted. For example,  the application of effective
(relative) alignment method in the $A\sim 140-150$ region of
superdeformation requires an accuracy in the prediction of relative
angular momenta on the level of $\sim 0.3\hbar$ and $\sim 0.5\hbar$ for
non-intruder and intruder orbitals, respectively
\cite{Rag.93,BHN.95,ALR.98}. In the highly deformed and SD nuclei from
the $A\sim 60-80$ mass region, these requirements for accuracy are
somewhat relaxed (see Refs.\ \cite{ARR.99,AF.05}). We expect
that in the $A\sim 130$ region, the relative alignments should be
predicted with a precision similar to that in  the $A\sim 140-150$
region. However, for a number of orbitals (for example, $\nu
[411]3/2^{\pm i}$, $\nu [532]5/2^{\pm i}$, $\nu 6_3^{-i}$, $\pi
[301]1/2^{+i}$, $\pi [413]5/2^{-i}$, $\pi [550]1/2^{-i}$),  the
calculated uncertainties in $j^{\text{eff}}_{\alpha}$ are close to
$0.5\hbar$, and this  probably prevents reliable assignments based on
the additivity principle for the configurations involving these
orbitals. The situation becomes even more uncertain if several orbitals
with high uncertainties in $j^{\text{eff}}_{\alpha}$ are occupied.

Let us also remark that while  theory provides effective alignments at a fixed
rotational frequency, relative alignments extracted from experimental data
may show appreciable frequency dependence (see for instance Ref.\ \cite{D.98}).
 Therefore, for
reliable configuration assignments,  measured relative alignments should be
compared with calculated ones  over a wide  frequency range.

\subsection{Variance and distribution of residuals}
\label{accuracy}

\begin{figure}
\centering
\includegraphics[width=8.0cm]{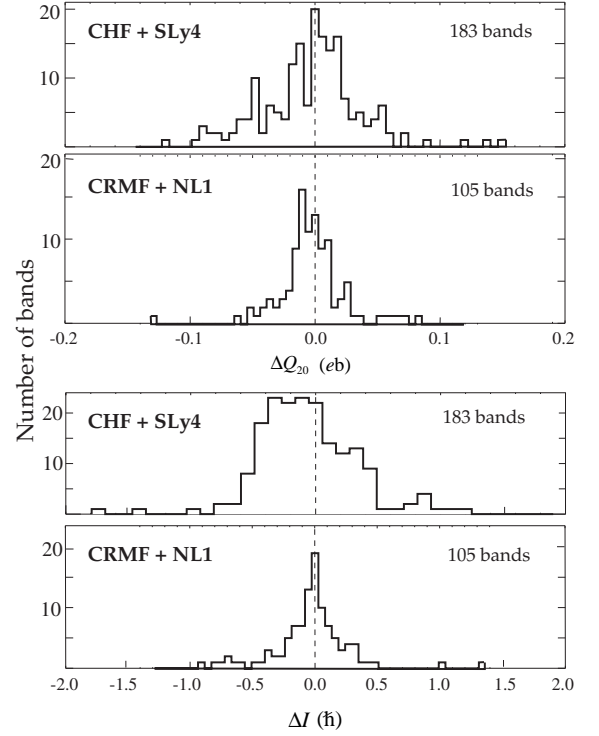}
\caption{Histogram of differences between  self-consistent values
obtained in CHF and CRMF
and those given by the additivity formula [see e.g.\ Eq.\
(\protect\ref{o1})]. The results for
$Q_{20}$ are shown in the two upper
panels and those for the total angular momentum  are displayed in the two
lower panels.
\label{deviat-q20}}
\end{figure}

\begin{figure}
\centering
\includegraphics[width=8.0cm]{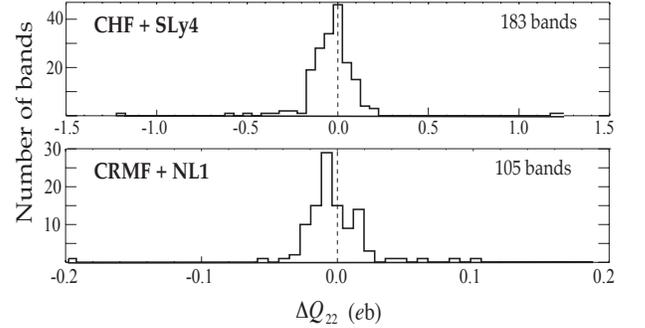}
\caption{Similar to Fig.\ \protect\ref{deviat-q20} except for $Q_{22}$.
\label{deviat-q22}}
\end{figure}

One of the main outcomes of this
study is the set of effective s.p.\  moments
$q_{20,\alpha}^{\text{eff}}$, $q_{22,\alpha}^{\text{eff}}$,
$q_{t,\alpha}^{\text{eff}}$, and alignments $j_\alpha^{\text{eff}}$. The quality
of the additivity principle   can be assessed by studying
the distribution of first moments of
residuals (\ref{residuals}), i.e., differences
 between the self-consistently calculated
values of physical observables and those obtained from the additivity
principle. For instance, for the quadrupole moment $Q_{20}$, the
quantity of interest is
\begin{eqnarray}
\Delta Q_{20}= \sum_{\alpha}^{} { c_{\alpha}(k) q_{20,\alpha}^{\text{eff}} }
  - \delta Q_{20} (k).
 \label{o1}
\end{eqnarray}
Deviations $\Delta Q_{22}$,  $\Delta Q_t$, and $\Delta J$
are given by similar expressions.
Figures \ref{deviat-q20} and \ref{deviat-q22} show distributions of these deviations. The
quality of the additivity principle for  $Q_{20}$ is shown in the top
two panels of Fig.\ \ref{deviat-q20}.
In the CHF model, the majority of  $\Delta Q_{20}$ values
(more than 97.8\% of the total number) fall comfortably within the interval
of $\pm$0.1\,eb. This corresponds to a relative distribution width of
about 1.3\%. In  CRMF,
the distribution is even narrower,
with more than 90\% of $\Delta Q_{20}$ values falling within
the $\pm$0.05\,eb interval, or less than 0.7\% of the total value.

The results for the total angular momentum  are shown in the bottom
panels of Fig.\ \ref{deviat-q20}. In  CRMF, the distribution of deviations
is very narrow, with only 10\% of the cases differing by more than
$\pm\hbar/2$. The  CHF histogram  is somewhat wider, but more than 90\%
of  deviations fall within the $\pm\hbar/2$ interval. Taking into
consideration that the experimental spins of highly deformed and SD
bands are often assigned with  uncertainties that are multiples of
$\hbar$, our results give considerable encouragement for theoretical
interpretations based on the method of relative (effective) alignments
\cite{Rag.93,BHN.95,ALR.98}.

In CHF and CRMF, the distributions of deviations of charge quadrupole
moments $Q_{22}$ (Fig.\ \ref{deviat-q22}) are relatively narrow. Again,
for CRMF, nearly 95\% of deviations fall within $\pm$0.025\,eb, and 98\%
fall within $\pm$0.1\,eb. For CHF, the distribution of deviations is
somewhat wider, with more than 90\% of deviations falling within the
$\pm$0.2\,eb interval.

We interpret these results as a strong indication that the additivity
principle works fairly well in self-consistent cranked theories. While
distributions of deviations in $Q_{20}$ and  $Q_{22}$ are rather similar
in  the CHF+SLy4 and CRMF+NL1 models (see top of Fig.\ \ref{deviat-q20} and
Fig.\ \ref{deviat-q22}), deviations in angular momentum differ between
these two approaches. Considering that  (i) the uncertainties in
$j_{\alpha}^{\text{eff}}$  are similar in both methods (Sec.\
\ref{Sect-jeff}), and (ii) shape polarization effects are not that
different (Sec.\ \ref{Sect-q22-qt}), one can conclude that  the
observed difference is due  to the polarization of time-odd mean fields.
However, the detailed investigation of this effect is beyond the scope
of this study.

\section{Conclusions}
\label{Concl}
 The  additivity principle in
highly  and SD rotational bands of the $A\sim 130$ mass
region has been studied within the
cranked Hartree+Fock theory based on the SLy4 energy density functional
 and the cranked
relativistic mean-field theory with the NL1
Lagrangian. The main results can be summarized as follows:
\begin{itemize}
\item
The two sets of effective s.p.\
charge quadrupole moments $q_{20}^{\text{eff}}$ and
$q_{22}^{\text{eff}}$, transition quadrupole moments
$q_t^{\text{eff}}$, and effective angular momenta $j^{\text{eff}}$
have been produced. This rich output
 allows for an easy and simple determination of
transition quadrupole moments $Q_t$
 in  highly deformed and SD bands  in the $A\sim 130$ mass  region. 
 In some cases, 
configuration assignments based
on the relative (effective) alignment method can 
 be done based on the
calculated values of effective angular momenta $j^{\text{eff}}$ (see, however, 
Sec.~\ref{Sect-jeff}).
\item
Our  statistical analysis of  distributions of residuals confirms that
the additivity principle is well fulfilled in the self-consistent
approaches that properly take into account polarization effects.
\item
The contribution from the triaxial degree of freedom to the transition
quadrupole moment is usually small, but it cannot be ignored when
aiming at a quantitative reproduction of experimental data. The average
magnitude
of $q_{22}^{\text{eff}}$ values is greater in the CHF+SLy4 model than in CRHF+NL1,
thus suggesting that the potential energy surfaces produced
in the former model are more $\gamma$-soft.
\item
For the majority of s.p.\  orbitals, there is a considerable
difference between the effective and bare expectation values
of one-body operators.
This indicates the importance of  polarization effects (shape polarization for
quadrupole moments and the shape and
time-odd-mean-field polarization  for
angular momentum alignment).
\item
With  very few exceptions, there is a great deal of
consistency between   CHF and CRMF results for the effective s.p.
moments and alignments.
\end{itemize}
So far, the additivity principle has been investigated only for
highly deformed or SD bands in the $A$$\sim$130-150
 mass region. It would be interesting to extend such studies to
other high spin structures. The most promising candidates are: (i)  terminating
bands in the $A\sim 110$ mass region characterized by very weak pairing
and appreciable  $\gamma$-softness \cite{AFLR.99},  and (ii)
SD rotational bands in the $A\sim 60$ and $A\sim 80$ mass
regions of superdeformation \cite{A6080}. Work along these lines is in progress.

\section{Acknowledgements}
Our study  was inspired by the experimental work of Laird {\em et al.}~\cite{A130-exp1}.
Stimulating discussions with Mark Riley are gratefully acknowledged.
The work was supported in part
by the U.S.\ Department of
Energy under Contract Nos.\ DE-FG02-96ER40963 (University of
Tennessee), DE-AC05-00OR22725 with UT-Battelle, LLC (Oak Ridge
National Laboratory), and DE-FG05-87ER40361 (Joint Institute for
Heavy Ion Research), and DE-FG02-07ER41459 (Mississippi State
     University); by the Latvian Scientific Council (grant No. 05.1724);
by the Polish Ministry of Science; 
by the  Academy of Finland and University of Jyv\"askyl\"a within the
FIDIPRO programme; and  by the European Union Social Fund and the research
program Pythagoras II - EPEAEK II, under project 80861.

\end{document}